\documentclass[apl,aps,twocolumn,showpacs,groupedaddress]{revtex4}
\usepackage[dvipdfm]{graphicx}

\begin{document}

\title{Non-volatile resistive switching in dielectric
superconductor YBa$_2$Cu$_3$O$_{7-\delta}$}

\author{C. Acha}
\thanks{Fellow of CONICET of Argentina}
\email{acha@df.uba.ar}
\affiliation{Departamento de F\'{\i}sica,
FCEyN, Universidad de Buenos Aires, Ciudad Universitaria, (C1428EHA)
Buenos Aires, Argentina}
\author{M. J. Rozenberg}
\affiliation{Departamento de F\'{\i}sica, FCEyN, Universidad de
Buenos Aires, Ciudad Universitaria, (C1428EHA) Buenos Aires,
Argentina}
\affiliation{Laboratoire de Physique des Solides,
CNRS-UMR8502, Universit\'e de Paris-Sud, Orsay 91405, France}

\date{\today}

\draft

\begin{abstract}
We report on the reversible, nonvolatile and polarity dependent
resistive switching between superconductor and insulator states at
the interfaces of a Au/YBa$_2$Cu$_3$O$_{7-\delta}$ (YBCO)/Au system.
We show that the superconducting state of YBCO in regions near the
electrodes can be reversibly removed and restored. The possible
origin of the switching effect may be the migration of oxygen or
metallic ions along the grain boundaries that control the intergrain
superconducting coupling. Four-wire bulk resistance measurements
reveal that the migration is not restricted to interfaces and
produce significant bulk effects.

\end{abstract}

\maketitle


The improvement of the performance of silicon based electronic
memories is expected to begin reaching its limits in a decade or
two. This is motivating a great deal of activity in the search for
alternative technologies. Among the most promising candidates one
finds the resistive switching (RS) effect in capacitor like
metal/transition-metal oxide/metal structures. It basically consists
of the sudden change on the conductance of the system upon a strong
electric voltage application on the electrodes. The key features are
that the changes are non-volatile and reversible. The RS effect has
already been reported in systems with a wide range of
transition-metal oxide dielectric. Most of the work has focused on
thin films \cite{Baikalov03,Choi06}, however, it is also observed
using bulk ceramic dielectrics \cite{Tsui04,Quintero05}. Up to now,
most of the work indicates that the effect occurs in regions near
the contact interfaces \cite{Chen05}, nevertheless, there is still
not a consensus on the physical origin of the switching mechanism
\cite{Hamaguchi06,Sato07,Quintero07}. \\

In this letter we demonstrate the RS effect in a system where the
dielectric is the cuprate perovskite YBa$_2$Cu$_3$O$_{7-\delta}$
(YBCO) which is a high critical temperature superconductor. We use a
ceramic YBCO which shows metallic resistivity and a superconducting
transition at $T_c \simeq 90~K$. This is in contrast to all previous
reports that used insulating or semiconducting dielectrics. Since
our main focus is to unveil the physical mechanism of the switching,
we consider a bulk dielectric where by means of a multi-electrode
configuration one can study the switching of the interfaces an bulk
behavior independently. After intense electric pulsing of a given
polarity, we find that one interface becomes less resistive showing
signatures of a superconducting transition at $T_c$. The other
becomes more resistive and semiconductive without showing any
anomaly at $T_c$. Interestingly, upon intense electric pulsing with
the opposite polarity, the behavior of the resistance of the
interfaces is interchanged. This switching effect is non-volatile
and implies that superconductivity can be both, suppressed  and
restored by electric pulsing.  In addition, using a four-wire
measurement, we find that the resistance of the bulk dielectric
beneath the pulsed electrodes is affected by the amount of pulsing
and that the effect is at least partially reversible. Therefore, the
pulsing induces a change in the oxide which may extend over hundreds
of microns from the contact interfaces into the bulk. We shall argue
that our observations suggest that strong electric pulsing may cause
migration of either oxygen or metallic ions along higher ion
mobility regions, such as grain boundaries of YBCO.

\begin{figure}[b]
\vspace{0mm}
\centerline{\includegraphics[angle=0,scale=0.30]{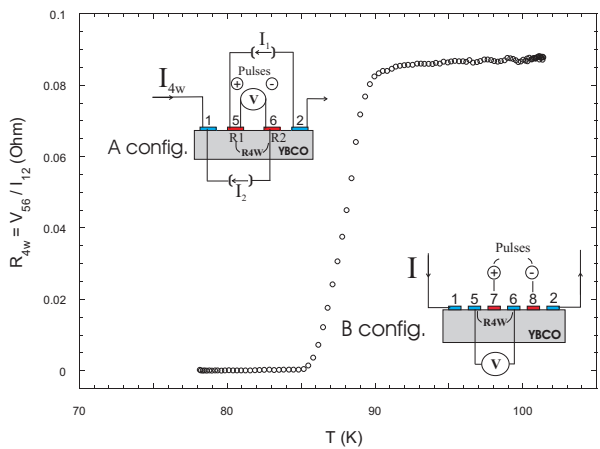}}
\vspace{0mm}\caption{(Color online) Temperature dependence of
$R_{4W}$ (V56/I12) of the YBCO sample before applying pulses ($T_c
\simeq 90~K$). The inset shows the two contact configurations
studied.} \vspace{-0mm} \label{fig:R4wycontactos}
\end{figure}

To study the RS effect we fabricated two configurations of contact
electrodes (A and B) on a ceramic YBCO ($J_c$(77 K) $\simeq 10^3$
A/cm$^2$). Samples were synthesized following a similar procedure to
the one described elsewhere \cite{Porcar97}. Contacts were made by
sputtering gold onto the entire width of one of the faces of a YBCO
slab (8x4x0.5 mm$^3$) and by using silver paint to fix the copper
leads. The mean electrode's width was 1 mm while their mean
separation was between 0.4 to 0.8 mm. The superconducting transition
of our sample and the contact configurations are shown in
Fig.~\ref{fig:R4wycontactos}. Square pulses of 10 V and 0.1 ms at 1
kHz were applied to electrodes 5-6 (7-8) of configuration A (B). The
maximum power applied during pulsing treatments was 0.25 W during 20
s, so that overheating is not expected to play a relevant role on
the observed effects. Resistance was measured using a standard DC
technique with a small positive and negative test current (10 to 100
$\mu$A). To measure a particular electrode's resistance, current was
forced to flow through that electrode and a third electrode, as
depicted in Fig.~\ref{fig:R4wycontactos} (configuration A). In this
way, the measured quantity is the contact resistance plus a bulk
YBCO contribution. The bulk resistance of the sample $R_{4W}$
(proportional to the sample resistivity) was measured independently,
using a standard four-wire technique (configuration A). The initial
resistance of contacts where in the range of 20 to 1000~$\Omega$,
while the bulk YBCO resistance was about 0.1~$\Omega$ for $T$ above
$T_c$. Thus the bulk contribution in configuration A was always
small or negligible.


\begin{figure}[h]
\vspace{-5mm}
\centerline{\includegraphics[angle=0,scale=0.3]{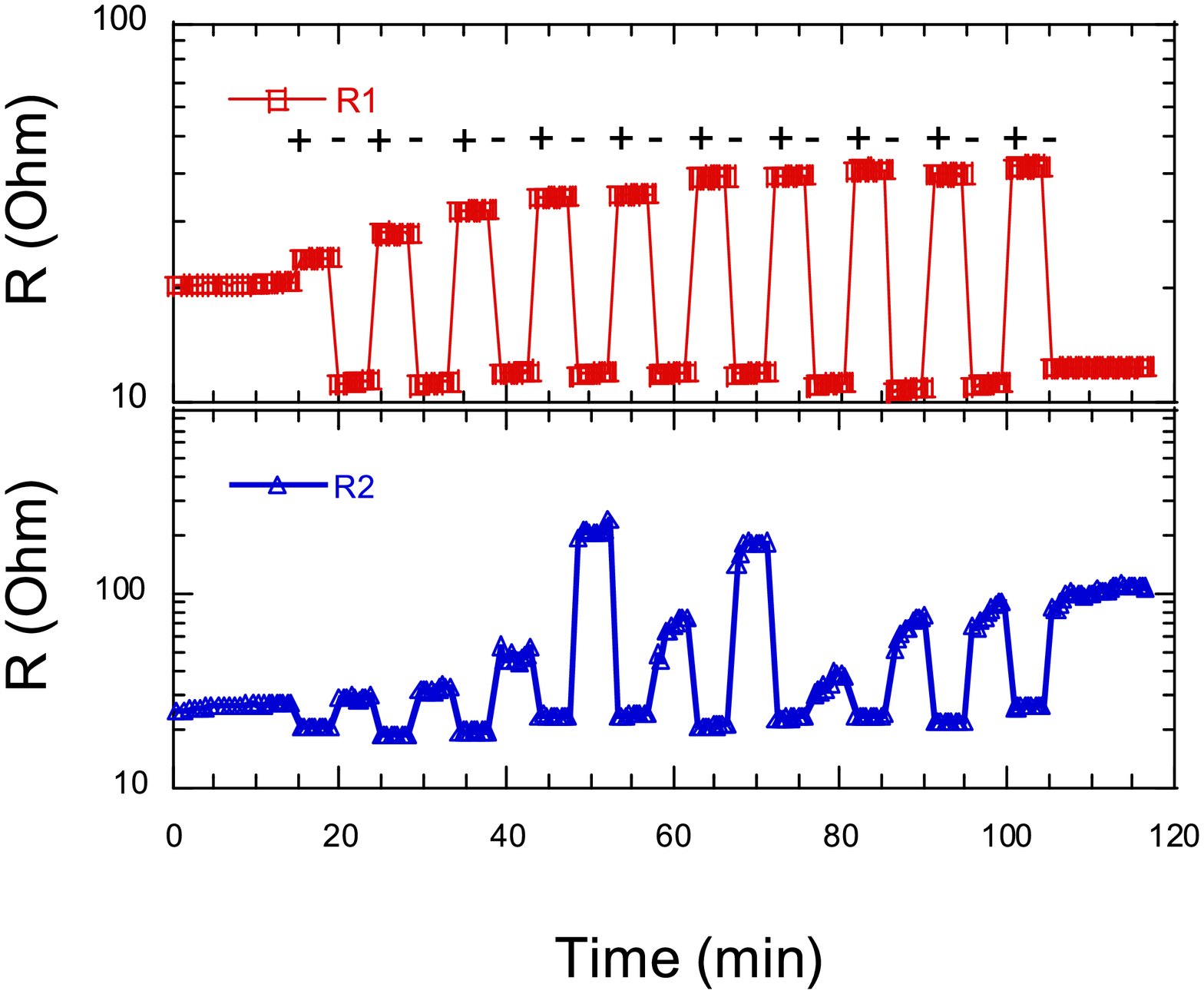}}
\vspace{0mm} \caption{(Color online) Room temperature time
dependence of $R_1$ ($V_{56}/I_{52}$) and $R_2$ ($V_{56}/I_{16}$)
while applying 2x10$^4$  ($+ = $+5;-6) and ($- = $-5;+6) electric
field pulses using the A contact configuration.} \vspace{0mm}
\label{fig:R1R2R4w_comp}
\end{figure}

Fig.~\ref{fig:R1R2R4w_comp} shows the time evolution at room
temperature of the contact resistance of the two pulsed electrodes
($R_1=V_{56}/I_{52}$ and $R_2=V_{56}/I_{16}$). The pulses were
applied in trains of 20,000 producing the large switching
transitions. The polarity applied to electrode 5 of the successive
trains is indicated in the figure. The resistance changes where
non-volatile and had a factor of up to 100 between the low and high
resistance states. A remarkable feature of the observed resistive
switching is that it always showed complementary behavior, ie, when
one interface increased its resistance the other one decreased it.
This type of behavior has also been observed in systems with
manganite dielectrics \cite{Tsui04,Quintero07}. We checked that the
instabilities and the slow relaxation observed during the
measurements were not due to sample heating. While in most instances
the cathode was associated with a lower resistance state, this
correlation was not always observed. \\

\begin{figure}[h]
\vspace{-5mm}
\centerline{\includegraphics[angle=0,scale=0.35]{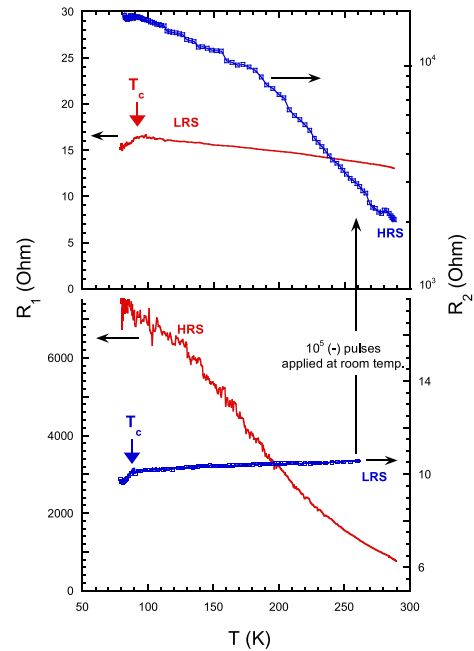}}
\vspace{0mm} \caption{(Color online) Temperature dependence of both
pulsed electrodes, before and after applying over $10^5$ negative
(-5; +6) pulses, using the A contact configuration at room
temperature. In the LR regime, the contact resistance is small and
tracks the R4w resistance, particularly its superconducting
transition.} \vspace{0mm} \label{fig:R1R2comp}
\end{figure}

\begin{figure}[h]
\vspace{0mm}
\centerline{\includegraphics[angle=0,scale=0.35]{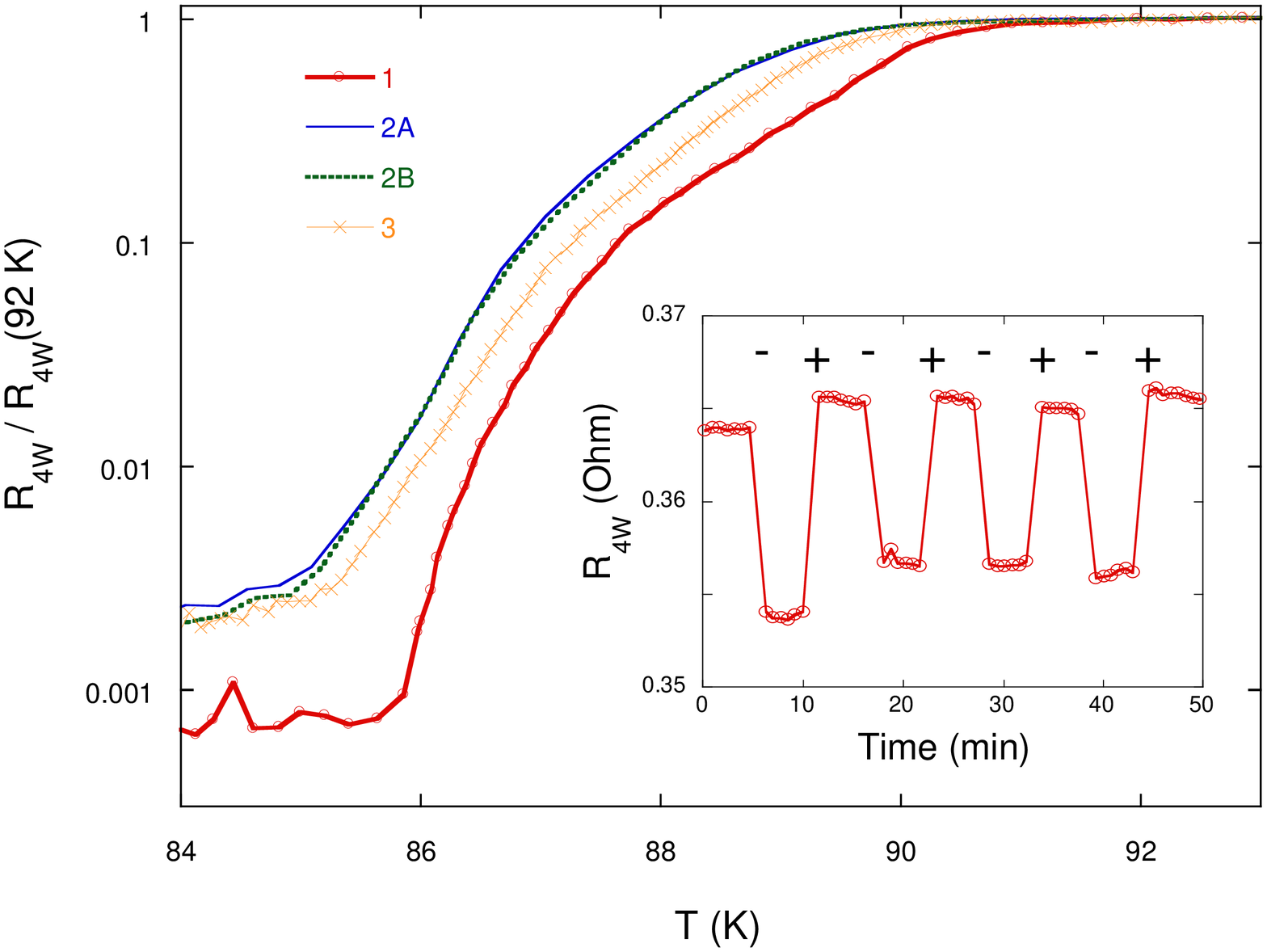}}
\vspace{-7mm} \caption{(Color online) Superconducting transition
sensitivity to pulses of opposite polarity, measured using the B
configuration. All the pulses were applied at room temperature. The
$T_c$ onset remains unchanged. $R_{4W}= V_{56}/I_{12}$; (1) initial
sample, (2A) $10^7$ $+$ cumulated pulses, width=100 $\mu$s; 1 kHz,
10 V between electrodes 7 (+) and 8 (-). (2B) is sample 2A measured
1 month later to check stability and repeatibility. (3) $10^8$ $-$
cumulated pulses. The inset shows the reversible behavior of
$R_{4W}$ upon applying trains of $2~10^4$ pulses of opposite
polarity at room temperature.} \vspace{0mm} \label{fig:R4wTc}
\end{figure}

To further characterize the switching mechanism, we investigated the
temperature dependence of the contact resistances. The results shown
in Fig.~\ref{fig:R1R2comp} where obtained by first applying a train
of pulses at room temperature that set one interface in the high
resistance state (HRS) and the other in the low resistance state
(LRS). The two contact resistances were measured simultaneously as
the system was cooled down. Then the sample is brought back to room
temperature, a new train of pulses of opposite polarity is applied
that produced the inversion of the resistive state of the contacts,
and the resistance was measured under a new cooling process. We
observed that if pulses were applied at low temperatures ($\sim$ 100
K), they produced substantially smaller effects. The resistance in
the HRS steeply increases with decreasing temperature, while in the
LRS it has a relatively weak temperature dependence. The most
interesting feature is that the LRS showed a signficant drop in the
resistance at $T_c$, while the HRS did not show any similar anomaly.

It is important to note that the observed drops in the contact
resistances $R_1$ and $R_2$ are about $ 1~\Omega$ (see the LRS
regime in Fig.~\ref{fig:R1R2comp}). Therefore, they cannot be
ascribed to a bulk change since the data of
Fig.~\ref{fig:R4wycontactos} demonstrates that contribution of the
bulk can be {\em at most} one order of magnitude smaller that the
observed drops. The implication of this observation is that pulsing
can actually suppress and {\em restore} the superconducting state of
the YBCO material in the neighborhood of the pulsed electrodes. \\

The natural questions that we now confront are: (i) what is the
physical mechanism that can suppress and restore the superconducting
state?, and (ii) what is the spatial extension of those regions?

Key insight on these issues was obtained from the study of the bulk
resistance $R_{4W}=V_{56}/I_{12}$ that involves a bulk region
section of the sample directly beneath electrode 7 in the contact
configuration B. More precisely, we focused on the influence of
electric pulsing on the bulk superconducting transition temperature
and lineshape. Results are shown in Fig.~\ref{fig:R4wTc}. We firstly
characterized the superconducting transition of the pristine sample.
Then, at room temperature, we applied a long train of $10^7$ (+)
polarity pulses on electrode $7$. Although local overheating can not
ruled out during this operation, only a small increase of
temperature was detected in the closest thermometer to sample. We
then measured again the superconducting transition and, within our
experimental resolution, we observed that the onset of
superconductivity $T_c$ remained unchanged while both, the
temperature width of the transition and the residual resistance
below $T_c$, exhibited significant increase. The measurement was
repeated on the stored sample one month later, showing negligible
relaxation, and, in addition, confirming the stability and the
thermal repeatability of our experimental setup. Upon application of
a reversed (-) polarity train of pulses at room temperature, the
effect on the superconducting transition was reversed. However, the
observed reversion was only partial, and with even up to $2.10^8$
(-) pulses we were not able to bring the sample back to the initial
condition. We checked that this was still the case also for a
successive change of the polarity of the electric excitation. On the
other hand, we did obtain a fully reversible effect for the bulk
resistance switching at room temperature, as is shown in the inset
of Fig.~\ref{fig:R4wTc}. These observations have several
implications. Firstly, the reversible control of the bulk resistance
suggest that pulsing affects the inter-grain Josephson junctions
coupling, supported by the observed suppression and restoration of
superconductivity without a change in its onset temperature value
$T_c$.\cite{Hilgenkamp02} Secondly, the effect on the coupling
between ceramic grains is also consistent with the observed increase
in the residual resistance, which is associated to a suppression of
superconducting percolating paths between measuring voltage
electrodes. Finally, the fact that R$_{4W}$ probes resistance
changes of regions of YBCO that are relatively far from the pulsed
electrodes ($\sim$1 mm), implies that pulsing may continue to
produce significant effects up to hundreds of microns away from the
pulsed contacts.

The peculiar effects of electric pulsing on the superconducting
state of YBCO revealed by our study provides clear fingerprints that
the underlaying physical mechanism for resistive switching  involves
the control of the inter-grain coupling. Most likely, this results
from migration of either oxygen or some metallic ion along the grain
boundaries, where their diffusion may be strongly enhanced by
electric fields. From the extensive studies done on cuprate
superconductors, it is well known that metallic-ion migration at
grain interfaces is one of the most effective parameters affecting
the superconducting properties of polycrystalline samples
\cite{Hilgenkamp02, Sydow99,Nava01}. On the other hand, the
diffusion of oxygen-ion may provoke the oxidation/reduction of the
inter-grain boundaries beneath the anode/cathode. From a more
general perspective, the inter-grain coupling control through ion
migration in grain boundaries is likely to be a common phenomenon to
all types of polycrystalline perovskites. Therefore, one may argue
that this mechanism may naturally account for the surprising
universality of the resistive switching effect in transition metal
oxides \cite{Quintero07,Szot06,Nian07}.

This work was partially supported by  UBACyT (X198), ANPCyT PICT
03-13517 and 02-11609, CONICET PIP 5609 and ECOS-Sud grants. We are
indebted to V. Bekeris and P. Levy for very fruitful discussions.

\break

\end{document}